\documentclass[aps,prl,10pt,superscriptaddress,twocolumn,floatfix]{revtex4-1}

\usepackage{natbib}
\usepackage[utf8]{inputenc}
\usepackage{xcolor}

\definecolor{linkblue}{RGB}{31,119,180}
\usepackage[
  unicode,
  colorlinks,
  citecolor=blue,
  linkcolor=linkblue,
  urlcolor=linkblue,
  bookmarks=true,
  bookmarksopen=true,
  bookmarksopenlevel=3,
  bookmarksnumbered=true]{hyperref}
\usepackage{graphicx}
\usepackage{amsmath}
\usepackage{amssymb}

\begin{document}

\title{Multi-step two-copy distillation of squeezed states via two photon subtraction}

\author{Stephan Grebien}
\affiliation{Institut f\"ur Laserphysik \& Zentrum f\"ur Optische Quantentechnologien, Universit\"at Hamburg, Luruper Chaussee 149, 22761 Hamburg, Germany}
\author{Julian G\"ottsch}
\affiliation{Institut f\"ur Laserphysik \& Zentrum f\"ur Optische Quantentechnologien, Universit\"at Hamburg, Luruper Chaussee 149, 22761 Hamburg, Germany}
\author{Boris Hage}
\affiliation{Institut f\"ur Physik, Universit\"at Rostock, 18051 Rostock, Germany}
\author{Jarom\'{i}r Fiur\'{a}\v{s}ek}
\affiliation{Department of Optics, {Faculty of Science}, Palack\'y University, 17. listopadu 12, 77900 Olomouc, Czech Republic}
\author{Roman~Schnabel}
\email{roman.schnabel@physnet.uni-hamburg.de}
\affiliation{Institut f\"ur Laserphysik \& Zentrum f\"ur Optische Quantentechnologien, Universit\"at Hamburg, Luruper Chaussee 149, 22761 Hamburg, Germany}

\date{\today}

\begin{abstract}
Squeezed states of light have been improving the sensitivity of gravitational-wave observatories and are nonclassical resources of quantum cryptography and envisioned photonic quantum computers. The higher the squeeze factor is, the higher is the quantum advantage. Almost all applications of squeezed light require multi-path optical interference, whose unavoidable imperfections introduce optical loss, degrade the squeeze factor, as well as the quantum advantage.
Here, for the first time, we experimentally demonstrate the distillation of Gaussian squeezed states that suffered from Gaussian photon loss. Our demonstration already involves two distillation steps. The first step improved the squeeze factor from 2.4\,dB to {2.8\,dB} by the subtraction of two photons. The second step improved the value from {2.8\,dB to 3.4\,dB} by a Gaussification protocol. It was realised on data measured at different times via an 8-port balanced homodyne detector and via data post-processing. The number of distillation steps can be increased by longer data sampling times, without additional hardware. We propose and discuss the application to quantum cryptography and photonic quantum computers.
\end{abstract}

\maketitle

\section{Introduction}
\vspace{-3mm}
Squeezed states \cite{Stoler1970,Lu1971,Yuen1976,Walls1983} represent one of the most important classes of nonclassical states of light. The applications of squeezed states cover a wide range of domains such as continuous-variable quantum teleportation \cite{Furusawa1998,Bowen2003}, one-sided device-independent quantum key distribution \cite{Gehring2015}, photonic quantum computing \cite{Larsen2019}, and last but not least the sensitivity enhancement of gravitational wave detectors \cite{LSC2011,Tse2019,Acernese2019}. Squeezed states of light can be deterministically generated in nonlinear optical media pumped with intense coherent laser beam \cite{Wu1986,Schnabel2017}. Squeezed states have Gaussian electromagnetic field uncertainties with variances  partly below those of the ground state.
It has been recognised that Gaussian squeezing is an irreducible resource \cite{Braunstein2005};  any combination of interference in passive linear interferometers, homodyne detection and feed-forward cannot distill from a squeezed-states ensemble of arbitrary size (a smaller) one with an enhanced squeeze factor \cite{Kraus2003}.
This no-go theorem is similar to that on entanglement distillation of Gaussian two-mode squeezed states with local Gaussian operations and
classical communication \cite{Eisert2002,Fiurasek2002,Giedke2002}.\\
Although significant squeeze factors can be produced \cite{Vahlbruch2008} --- with quantum uncertainty variances up to a factor of 32 (15\,dB) below ground state variance \cite{Vahlbruch2016} --- optical loss in down-stream applications reduces the factual squeeze factor while keeping the states' Gaussian character.
It is therefore of fundamental interest to make distillation of Gaussian squeezing possible. 
In the past, passive Gaussian operations were used to distill special kinds of non-Gaussian squeezed states, employing protocols that cannot counteract the typical Gaussian optical loss \cite{Heersink2006,Franzen2006}, and conditioning on photon subtraction 
was employed to increase the entanglement and squeezing of two-mode squeezed vacuum states \cite{Ourjoumtsev2007,Takahashi2010,Kurochkin2014,Dirmeier2020}.

\begin{figure}[b!]
\vspace{-4mm}
\centerline{\includegraphics[width=0.86\linewidth]{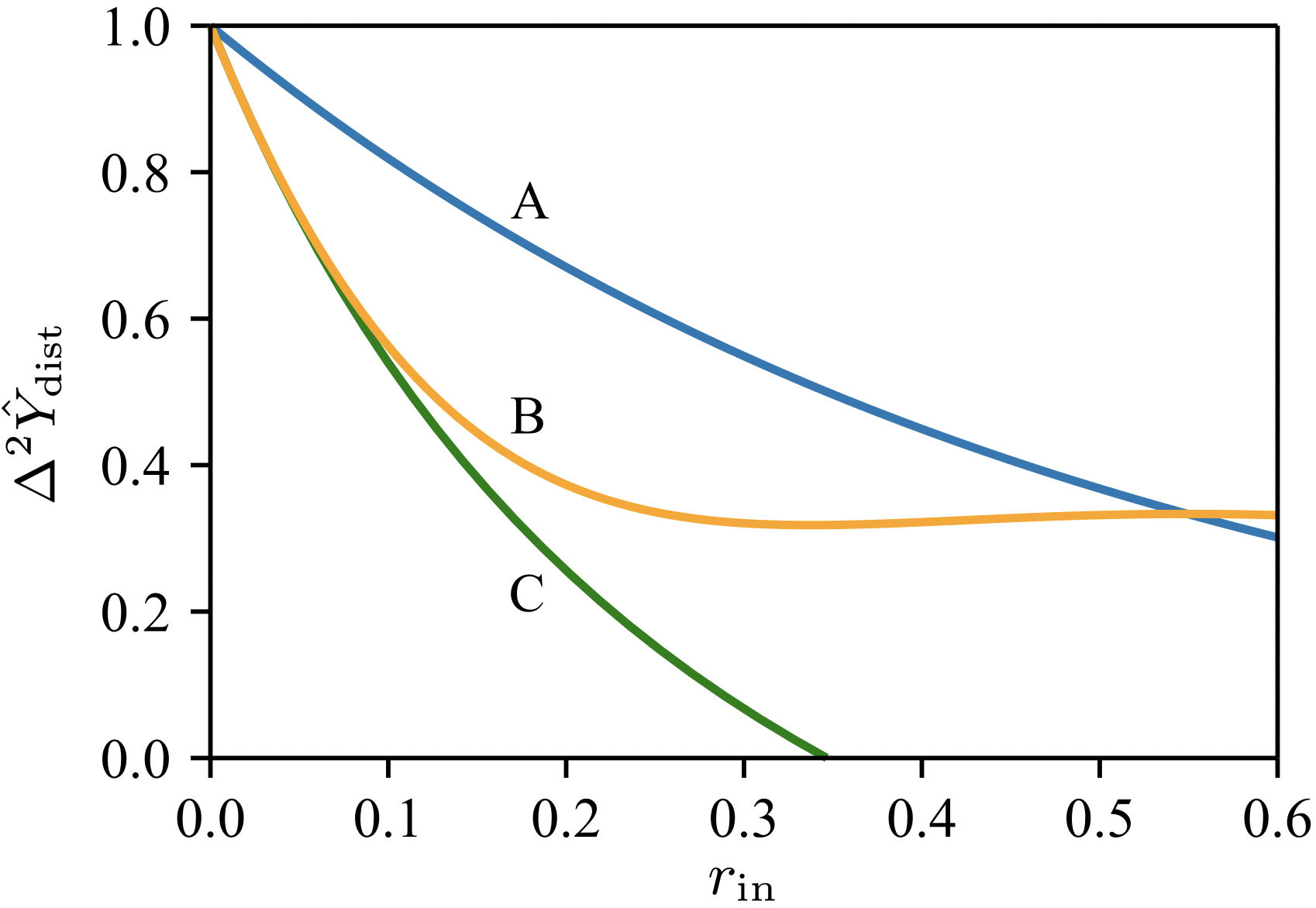}} 
\vspace{-2mm}
\caption{{\bf Potential of our distillation protocol} -- Variance of distilled squeezed quadrature uncertainty {$\Delta^2 \hat Y_{\rm dist}$} as a function of the initial squeeze parameter $r_{\rm in}$ before distillation. A: Initial, pure squeezed vacuum state. B: Two-photon subtracted squeezed vacuum state. C: Asymptotic limit of multi-step squeezing distillation. For an initially 3\,dB-squeezed state ($r_{\rm in} \approx 0.3466$), the squeeze factor can be increased to infinity ($\Delta^2 \hat Y_{\rm dist} \rightarrow 0$).
}
\label{fig:1}
\vspace{-3mm}
\end{figure}

Here, we demonstrate for the first time the distillation of Gaussian continuous-wave single-mode squeezed states. 
Our approach combines conditioning on two-photon subtraction with a second distillation step that further improves the squeeze factor and also Gaussifies the state. Our protocol represents a breakthrough because the second step can be repeated arbitrary many times depending on the amount of sampled data without the requirement for upscaling the hardware resources. Our analysis shows that arbitrary high squeeze factors are possible in theory. We make our protocol possible simultaneously sampling the non-commuting $X$ and $Y$ quadratures of the output field of a single continuous-wave squeezing resonator. Having simultaneously recorded $X$ and $Y$ for every individual mode, the distillation and Gaussification steps are done by probabilistic data post-processing. Our proof-of-principle distillation experiment enhances the squeeze factor over the relevant 3\,dB threshold, which for instance allows to surpass the no-cloning limit in quantum teleportation \cite{Bowen2003}.

\section{Quantum theoretical aspects of squeezed state distillation}
\vspace{-2mm}

Let us consider a pure single-mode squeezed vacuum state as a canonical example. Such a state has Gaussian uncertainties and can be written as a superposition of number states (Fock states) as
\begin{equation}
|\psi(r)\rangle=\frac{1}{\sqrt{\cosh(r)}}\sum_{n=0}^\infty (\tanh r)^n\frac{\sqrt{(2n)!}}{2^n n!}|2n\rangle \,
\label{eq:squeezedvacuum}
\end{equation}
where $r$ is the squeeze parameter \cite{Stoler1970}, and the variances of anti-squeezed and squeezed quadratures normalised to the vacuum uncertainty read {$\Delta \hat X^2 = e^{2r}$ and $\Delta \hat Y = e^{-2r}$}, respectively. The Heisenberg uncertainty relation for this normalisation reads $\Delta \hat X^2 \cdot \Delta \hat Y^2 \ge 1$.
The photon number distribution of the pure squeezed vacuum state exhibits the famous even-odd oscillations \cite{Lu1971} and only even Fock states are present in the expansion\,(\ref{eq:squeezedvacuum}). 

The squeezing of the state can be enhanced by keeping only some states, conditioned on the successful subtraction of two photons. 
The subtraction of two photons preserves the structure of the state in Fock basis and it enhances the amplitude of the two-photon state with respect to the amplitude of the vacuum state. 
The (non-normalised) state after subtraction of two photons $|\psi_{2S}(r)\rangle=\hat{a}^2|\psi(r)\rangle$ reads
\begin{equation}
|\psi_{2S}(r)\rangle= \frac{\tanh r}{\sqrt{\cosh(r)}}\sum_{n=0}^\infty (2n+1) (\tanh r)^n\frac{\sqrt{(2n)!}}{2^n n!}|2n\rangle.
\label{psi2S}
\end{equation}
This state is non-Gaussian and well approximates an `even' Schr\"odinger-cat-like state formed by the superposition of two displaced coherent (squeezed) states \cite{Takahashi2008,Marek2008}.
The variances of the squeezed and anti-squeezed quadratures of the non-Gaussian state in Eq.\,(\ref{psi2S}) can be analytically expressed,
\begin{eqnarray}
\Delta \hat X^2 &=& e^{2r} \!\left[1+4\frac{\sinh r\cosh r+2\sinh^2 r}{2\sinh^2r+\cosh^2r}\right], \nonumber \\[2mm]
\Delta \hat Y^2 &=& \, e^{-2r} \: \left[1-4\frac{\sinh r\cosh r-2\sinh^2 r}{2\sinh^2r+\cosh^2r}\right] \, . 
\end{eqnarray}
We find that the subtraction of two photons enhances the squeezing if, and only if $\tanh r <1/2$ ($\approx 4.8$\,dB), i.e.~only for moderate squeeze factors. We found out, however, that subsequent Gaussification of such states can in principle improve the squeeze factor arbitrarily.
Since the two-photon-subtracted state $|\psi_{\rm 2s}(r)\rangle$ is non-Gaussian, its squeezing can be further enhanced by an iterative Gaussification procedure, where two copies of the state are combined at a balanced beam splitter, and one output is accepted if the other output mode is projected onto vacuum. The output then forms the input for another iteration of the Gaussification \cite{Browne2003,Eisert2004}. The squeeze parameter $r_{\rm 2sG}$ of an asymptotic Gaussian state obtained by iterative Gaussification of the state $|\psi_{\rm 2s}(r)\rangle$ is given by
\[
\tanh r_{\rm 2sG} = 3\tanh r.
\]
The Gaussification converges only for $\tanh r<1/3$ (less than 3\,dB squeezing), and when $\tanh r \rightarrow 1/3$, arbitrarily strong squeezing can be distilled in principle from the initially close to 3\,dB-squeezed state.
This is illustrated in Fig.\,\ref{fig:1}, where we plot the squeezing variance of the initial squeezed vacuum state, the two-photon subtracted state, and the asymptotic Gaussified state. Even better performance can be obtained if the photon subtraction is combined with coherent displacement. 
A modified two-photon subtraction operation  $\hat{a}^2-\delta^2$ with suitably chosen coherent amplitude $\delta$ can enhance the squeezing of any single-mode squeezed vacuum state $|\psi(r)\rangle$, and an arbitrary strong squeezing can be asymptotically distilled from any $|\psi(r)\rangle$. 
See Supplementary Material for details.

\begin{figure}[t!!!!!!!!]
\centerline{\includegraphics[width=0.91\linewidth]{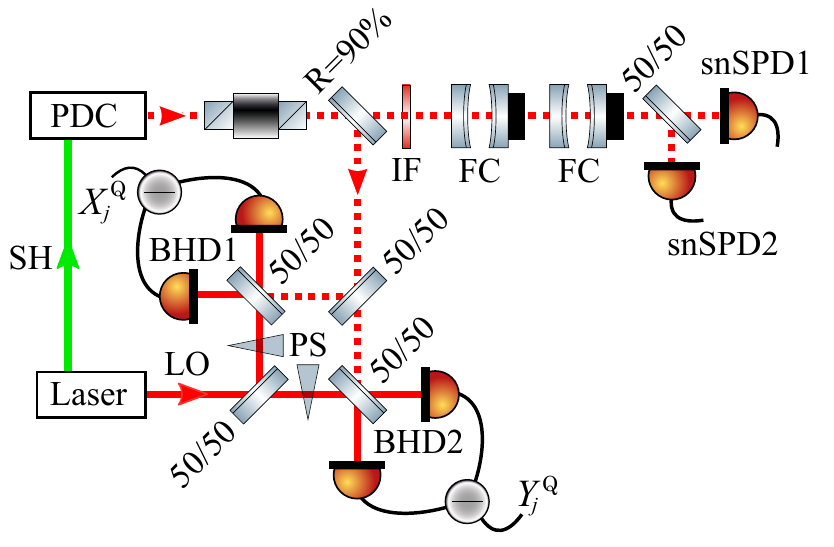}}
\vspace{-1mm}
\caption{
{\bf Optical setup} -- Resonator-enhanced parametric down-conversion (PDC) produced a beam of subsequent modes in identical squeezed vacuum states. 10\% of the states' energy was tapped and distributed onto two superconducting nanowire single-photon detectors (snSPD1,2). 90\% of the optical energy was also split and absorbed by two balanced homodyne detectors (BHD1,2) that simultaneously measured values of the non-commuting quadratures $\hat X^Q$ and $\hat Y^Q$, establishing a so-called 8-port BHD. The subscript `$Q$' indicates data taken on halves of the beam. An interference filter (IF) and two optical filter cavities (FC) rejected the optical spectrum outside the BHD bandwidth. LO: continuous-wave local oscillator (1064\,nm), PS: phase shifter, SH: second-harmonic pump field (532\,nm).}
\label{fig:2}
\end{figure}
\begin{figure}[t!!!!!!!!]
\vspace{0mm}
\includegraphics[width=0.88\linewidth]{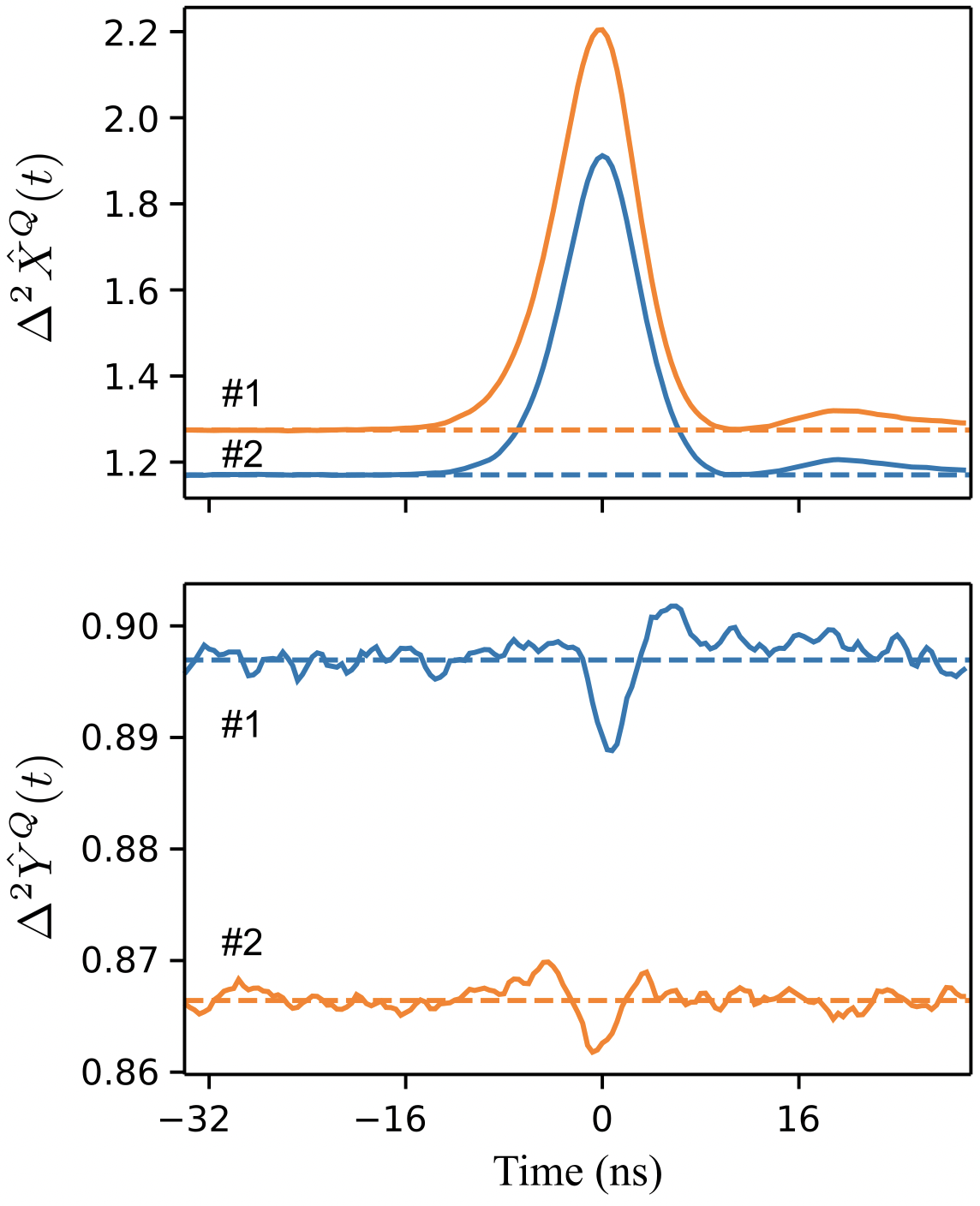}
\vspace{-1mm}
\caption{
{\bf Variances after two subtracted photons} -- Shown are the results of two independent measurement runs with slightly different initial squeeze factors. The quadrature variances $\Delta^2 \hat X^Q (t)$ (top, from BHD1) and $\Delta^2 \hat Y^Q (t)$ (bottom, from BHD2) include the time when both snSPDs clicked, to which both x-axes are referenced to ($t=0$). The traces are calculated from $10^6$ individual measurements on halves of the beam and represent data from which the Husimi Q-function \cite{Husimi1940} can be calculated. Around $t=0$, the anti-squeezing as well the squeezing are enhanced, which represents the distillation success of the first step of our protocol. 
{Note that the data includes frequencies outside the bandwidth of the squeezing resonator. This dilutes the actual squeeze factor from $\,\Delta^2 \hat Y^Q (t) \approx 0.79$ ($2.4$\,dB) to about 0.866 in $\# 2$.}
}
\label{fig:3}
\end{figure}
\begin{figure*}[t!!!!!!!!!!!!!!!!!!!!!!!!]
\vspace{0mm}
\includegraphics[width=0.95\linewidth]{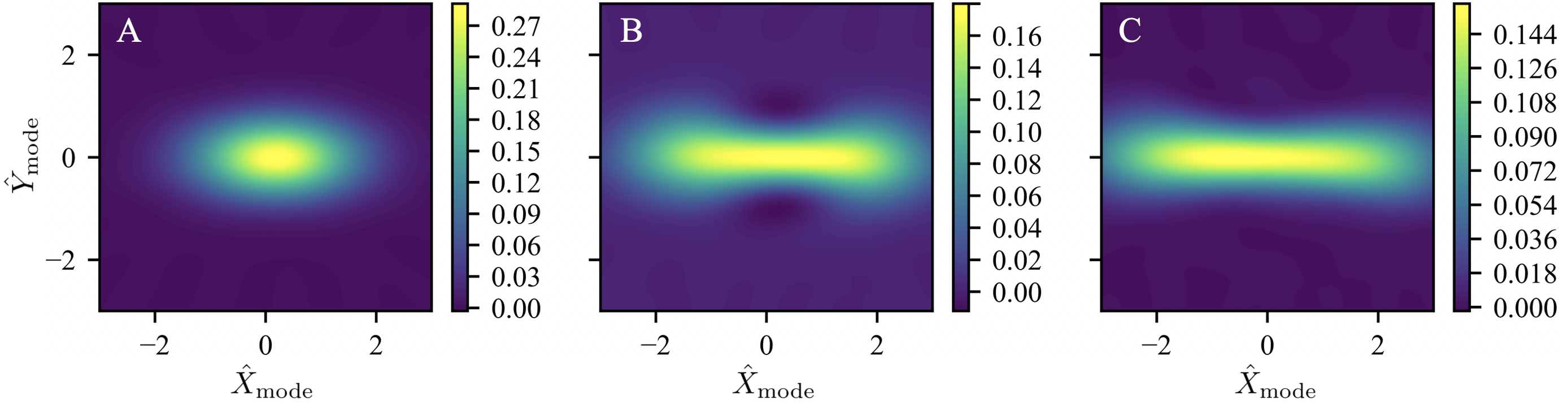}
\vspace{-1mm}
\caption{
{\bf Reconstructed Wigner functions} -- A: The initial 2.4\,dB-squeezed vacuum state. B: {The initial state distilled by the subtraction of two photons yielding 2.8\,dB squeezing. C: Example of the subsequently two-copy-two-step-distilled and Gaussified squeezed vacuum state for $\bar{n}=1.3$ having 3.14\,dB squeezing. The Gaussification step reduced the ensemble size to the fraction $P_{\mathrm{svv}}=0.246$. Additional two-copy distillation steps are possible in principle if the amount of samples is sufficiently high. The `mode' is defined by the temporal shape $f(t)$ and its Fourier transform limited spectrum.}
}
\vspace{2mm}
\label{fig:4}
\end{figure*}

\section{Experimental}
\vspace{-2mm}
Fig.\,\ref{fig:2} shows the schematic of the optical setup. The master laser was a continuous-wave Nd:YAG laser that provided an ultra-stable light beam of up to 2\,W at 1064\,nm in a TEM{\small 00} mode. Most of this light was frequency doubled and used to pump a resonator-enhanced, type\,I degenerate parametric down-conversion (PDC) process below oscillation threshold. The nonlinear material inside the resonator was periodically poled KTiOPO$_4$ (ppKTP). The PDC resonator produced a continuous stream of squeezed vacuum states in a TEM{\small 00} beam with a squeeze factor of up to $e^{2r}=10$ (10 dB). For the experiments here, however, we reduced the pump power to produce states with a squeeze factor of about 2 (3\,dB) and consequently very high purity \cite{Vahlbruch2016}. This squeezed vacuum beam was split with a power ratio of 10/90. The `signal' beam (higher fraction) was measured with a pair of balanced homodyne detectors (BHD) with quantum efficiencies above 98\%. The balanced homodyne detectors were arranged in the so-called `8-port' configuration, i.e.~one BHD continuously measured the squeezed quadrature and the other one simultaneously the anti-squeezed quadrature on halves of the beam. 

The `trigger' beam (10\% fraction) was spectrally filtered by an interference filter with a transmission peak at 1064\,nm and a HWHM of 0.6\,nm and subsequently by two length-controlled Fabry-Perot resonators \cite{Neergaard-Nielsen2006}. 
The purpose of these filters was to remove all squeezed field components that were outside the detection bandwidth of the BHDs. The filtered beam was split and measured by two super-conducting nanowire single-photon detectors with quantum efficiencies greater than 93\% (snSPD1,2). The data of the BHDs was only analysed, when both of the snSPDs detected a photon. In this case, the signal beam contained a mode in a squeezed vacuum state of which 2 photons were subtracted.

\section{Results and discussion}
\vspace{-2mm}
We ran the entire distillation protocol two times, on two different days. In both runs, we recorded $1.65\times 10^6$ two-photon subtraction events. For each event $j$ we have simultaneously sampled time-resolved quadrature values $\hat{X}^Q_j(t)$ and $\hat{Y}^Q_j(t)$ within a $64$~ns  long time window centred on the subtraction event. 
Fig.\,\ref{fig:3} shows the two pairs of time-resolved variances $\Delta^2 \hat X^Q(t)=(\Delta^2 \hat X(t)+1)/2$ and $\Delta^2 \hat Y^Q(t)=(\Delta^2 \hat Y(t)+1)/2$, where $\Delta^2 \hat{X}(t)$ and $\Delta^2 \hat{Y}(t)$ denote the corresponding variances of the signal beam before it was split for 8-port balanced homodyning. (The factor of $1/2$ is due to the vacuum uncertainty entering the open port.)
At the times around successful two-photon subtraction, which we deliberately set to zero in Fig.\,\ref{fig:3}, the anti-squeezed variance increased (upper plot) and the squeezed variance got reduced (lower plot). The observed dip represents the direct experimental manifestation of squeezing enhancement via two-photon subtraction. Variances sufficiently far away from the time of photon subtraction can be approximated by horizontal lines, which represent the levels of anti-squeezing and squeezing without photon subtraction, respectively.

The temporal shape $f(t)$ of the mode that contained the two-photon subtracted state was extracted from the temporal covariance matrix of the anti-squeezed quadrature \cite{Morin2013}. We took into account the non-trivial structure of the covariance matrix for vacuum input which is related to the response function of our detector, see Supplementary Material for details. The recorded windows were weighted by $f(t)$ and integrated over time. The result were $1.65\cdot10^6$ quadrature pairs $\hat X^Q_{\rm mode}$ and $\hat Y^Q_{\rm mode}$ for each of the two runs, representing results of 8-port homodyne detection on identical modes with the temporal profile $f(t)$. The same procedure was applied to characterise the same mode in a vacuum state to yield the quadrature variances for shot-noise normalisation. The two-dimensional histogram of the measurement outcomes $\beta=X^Q_{\mathrm{mode}}+iY^Q_{\mathrm{mode}}$ corresponded to the Husimi $Q$-function, which represented its complete information. From this, we reconstructed the density matrix in Fock basis using the statistically motivated and robust maximum-likelihood reconstruction algorithm.

Fig.\,\ref{fig:4} shows Wigner functions \cite{Wigner1932} of states of the mode with temporal profile $f(t)$ 
calculated from the reconstructed density matrices. It represents the result of our work.
Panel {A} shows the initial Gaussian squeezed vacuum state {before photon subtraction with a squeeze factor of 2.4\,dB ($10\!\cdot\!{\rm log}_{10} \Delta^2 \hat Y_{\rm mode}$). 
Panel {B} shows the 2-photon subtracted state of the same mode. The squeeze factor increased to 2.8\,dB.
We also determined the quadrature variances directly from the measured quadratures and found excellent agreement. 
The state in panel {B} is clearly non-Gaussian shape, which is a necessary condition for a second distillation step without further photon subtraction.
Since we used 8-port balanced homodyne detection, the full phase space data was recorded for any individual copy. This enabled us to emulate the interference of two copies of the state at a balanced beam splitter via data post-processing. This approach to 2-copy distillation is as efficient as the hardware-based version with in fact {\it perfect} quantum memories at hand \cite{Abdelkhalek2016}.
We post-processed pairs of measurement outcomes $\beta_{2j}$ and $\beta_{2j+1}$, where $1\leq j \leq5\times 10^5$, and,
\[
\beta_{j+}=\frac{1}{\sqrt{2}}(\beta_{2j}+\beta_{2j+1}), \qquad \beta_{j-}=\frac{1}{\sqrt{2}}(\beta_{2j}-\beta_{2j+1}).
\]
If the amplitude in the constructively interfering output by chance obeys $|\beta_{+}|^2 < \bar{n}$, where $\bar{n}$ is a freely choosable hard boundary, a state with improved squeezing emerges in the destructively interfering output port. The complex amplitude $\beta_{j-}$ represents further distilled, partly gaussified states. Fig.\,4\,{C} shows the result after one such additional Gaussification step with $\bar{n}=1.3$. The squeeze factor is increased to 3.14\,dB. 

Fig.\,\ref{fig:5} shows the improvement of the anti-squeezed variance $\Delta^2 \hat{X}$ (top) and the squeezed variance $\Delta^2 \hat{Y}$ (bottom) versus the success probability $P_{\mathrm{svv}}$ (the survival rate) of our Gaussification protocol. Since two input copies produce only one output copy, we have $P_{\mathrm{svv}} \leq 0.5$, and the lower the threshold $\bar{n}$ is, the lower the success probability. 
The improvement in squeeze factor (bottom) directly depends on the survival rate. The distilled squeezing reaches up to 3.4\,dB. For $25$\% survival rate it already exceeds $3.1$\,dB.  
Convergence of iterative Gaussification with finite acceptance window can be analysed for Gaussian acceptance probability $P(\beta)=\exp(-|\beta|^2/\bar{n})$ by invoking the formalism of characteristic functions of non-Hermitian operators \cite{Campbell2012} or, more straightforwardly, by interpreting this protocol 
as a combination of a lossy channel with transmittance $T=1/(\bar{n}+1)$ and the original Gaussification scheme with projection onto vacuum, see Supplementary material.  For the state in Fig.\,\ref{fig:4}\,B we find that the iterative Gaussification converges for $\bar{n} > 0.3$. 

\begin{figure}[h!!!!!!!]
\vspace{0mm}
\includegraphics[width=0.77\linewidth]{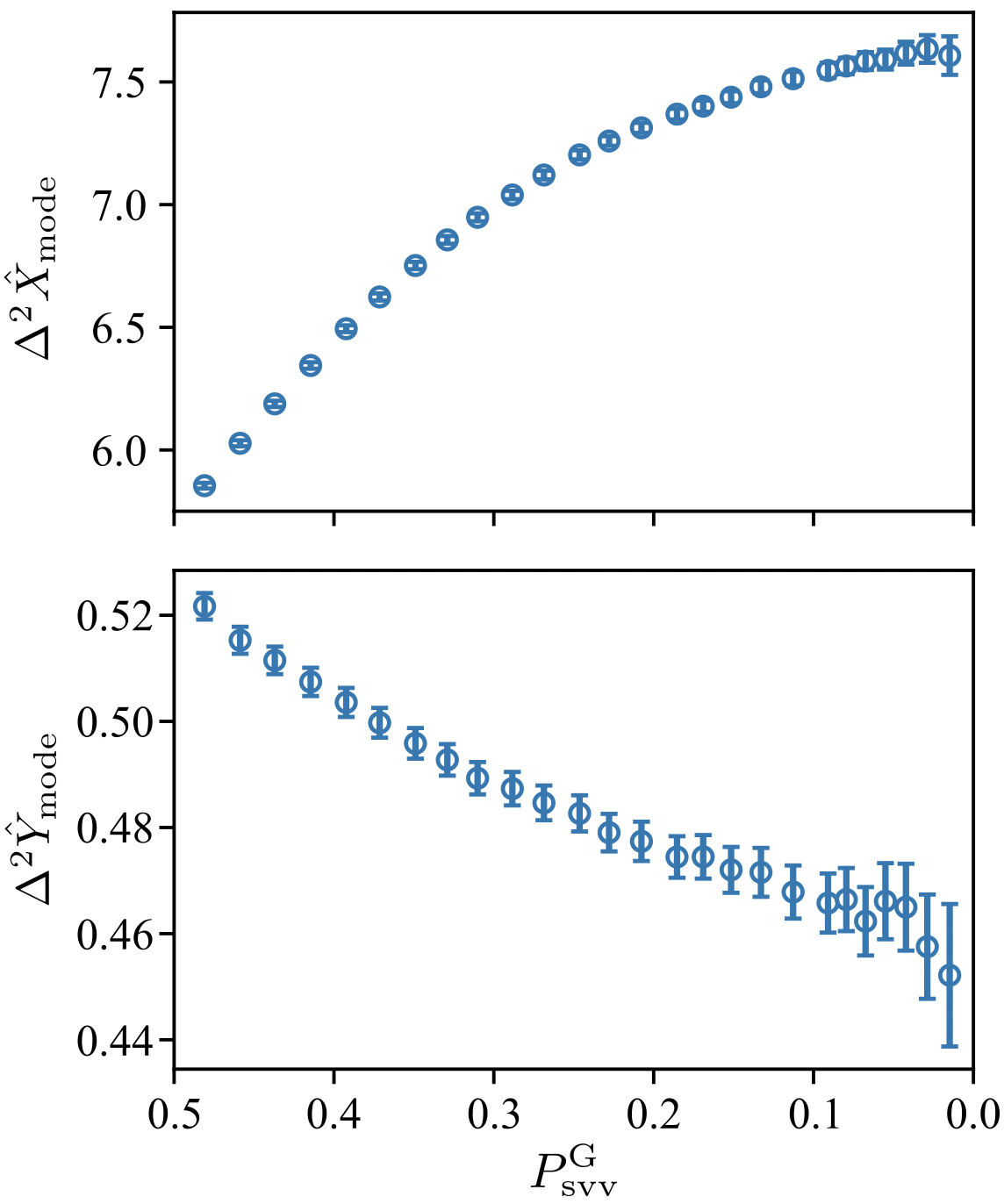}
\vspace{-3mm}
\caption{
{\bf Success rate of the Gaussification} -- $P^{\rm G}_{\rm svv}$ versus the distilled squeezed and anti-squeezed variances as in Fig.\,\ref{fig:4}\,\textbf{C}. For a high success probability, the increase in squeezing is marginal, while for lower success probability the squeezing increases significantly while greatly reducing the ensemble size.
}
\label{fig:5}
\vspace{-1mm}
\end{figure}

Even the distillation of pure squeezed vacuum states from initially mixed states is possible, however requires a more challenging non-Gaussian operation. We find that pure state distillation is possible with a Fock-state filter $\hat{F}_1=\hat{n}-1$
that completely eliminates the single-photon term in the density matrix. The quantum filter $\hat{F}_1$ can be realised with a single-photon catalysis \cite{Ulanov2015,Lvovsky2002}, which requires an ancilla single photon state that interferes with the signal at a suitably unbalanced beam splitter and the success is heralded by detection of exactly one photon at the ancilla output port. Alternatively, the operation $\hat{F}_1$ could be realised by a coherent combination of single-photon addition and subtraction \cite{Parigi2007,Costanzo2017}. 
For any input mixed state, the filtered density matrix $\hat{\rho}^{F}=\hat{F}_1 \hat{\rho}\hat{F}_1^\dagger$ 
will have vanishing density matrix elements $\rho_{0,1}^F$, $\rho_{1,0}^F$ and $\rho_{1,1}^F$. The theory of iterative Gaussification procedure then predicts that the Gaussification will converge to a pure  Gaussian squeezed vacuum  state whose squeeze parameter $r$ is completely determined by the parameter 
$\sigma_{2,0}^F=\rho^F_{2,0}/\rho^F_{0,0}$, namely $\tanh r=\sqrt{2}|\sigma_{2,0}^F|$. Remarkably, with the non-Gaussian operation $\hat{F}_1$ we can extract squeezing even from initial classical states such as coherent states. The necessary and sufficient requirement is that the initial state has non-vanishing coherence $\rho_{2,0}$ between the vacuum and two-photon Fock states and $|\sigma_{2,0}^F|<1/\sqrt{2}$.

\section{Conclusion}
\vspace{-2mm}
Ensembles of squeezed states with Gaussian quantum uncertainties can only be distilled to higher squeeze factors by non-Gaussian means.
Firstly, we provided the experimental proof-of-principle that mixed Gaussian squeezed states can be distilled to an increased squeeze factor by the subtraction of two photons. The distilled states were an ensemble of non-Gaussian squeezed states. In a second step, we successfully demonstrated two-copy distillation including Gaussification. For this, we realised an innovative approach that requires only a single beam of light. Two balanced homodyne detectors measured simultaneously the two non-commuting squeezed and anti-squeezed field observables on two halves of the beam. This approach not only provides the full information about the ensemble, but also individual-mode information for emulating two-copy distillation and Gaussification via data post-processing. Our approach is as efficient as using two simultaneously produced squeezed states plus perfect quantum memories \cite{Abdelkhalek2016}.\\ 
Photon events from Gaussian states are always probabilistic and thus the number of distilled states always lower than that of the input states.
For this reason, squeezed state distillation is most likely not useful in sensing {where signals with unknown shapes occur during finite time intervals. 
Quantum key distribution and quantum computing protocols are different, because they can be repeated many times with the same setting. The result represents a measured ensemble to that our approach is applicable in principle. The potential of our approach for one-sided device-independent quantum key distribution and measurement-based photonic quantum computing, however, requires further theoretical research.}\\


\textbf{Methods} ---
\emph{Data sampling:} 
The continuous stream of analogue output voltages of the two BHDs were recorded by a fast 2-channel data acquisition card (DAQ). Each channel was sampled with $2.5 \cdot 10^9\!$/s, with 12 bit resolution and a peak to peak range of 500\,mV. The DAQ-card featured a fast trigger channel, having a sampling rate of $2 \cdot 10^{10}\!$/s and a slower synchronisation (SYNC) channel, running with a sampling rate of $312.5 \cdot 10^6\!$/s in combination with a temporal pattern generator for activating and deactivating the trigger channel. We used the trigger and SYNC channels in the following way. 
The trigger channel was always deactivated by a ``1'' signal from the pattern generator. Only when a rising edge event from snSPD1 arrived at the SYNC channel, the pattern generator activated the trigger channel for 6.4\,ns by sending a ``0'' signal with a short time delay. The delay was compensated by a slightly longer cable that connected snSPD2 and the trigger channel. 
When a rising edge event was detected on the (activated) trigger channel, 160 quadrature pairs $\hat{X}^Q(t_j)$ and $\hat{Y}^Q(t_j)$ were sampled within a 64\,ns window. With this procedure, we reached an event rate of about one hundred 2-photon-subtracted states per second.

\vspace{12mm}
\begin{acknowledgments}
\textbf{Acknowledgments} ---
This work was funded by the Deutsche Forschungsgemeinschaft (DFG, German Research Foundation) -- SCHN 757/7-1. 
J.F.  acknowledges support by the Czech Science Foundation under Grant No. 21-23120S.\\
\end{acknowledgments}

\textbf{Author Contributions} ---
J.G., J.F. and R.S. planned the experiment. S.G., J.G., and B.H. built and performed the experiment. S.G. and J.F. provided the theoretical analysis. S.G., J.F., and R.S. prepared the manuscript.\\

\textbf{Data availability} ---
The data that support the plots within this paper and other findings of this study are available from the corresponding author upon reasonable request.\\

\textbf{Competing interests} ---
The authors declare no competing interests.\\

\textbf{Additional information} ---
Correspondence and requests for materials should be addressed to R.S.\\


~\\
~\\

\large{{\bf SUPPLEMENTARY MATERIAL}}

\section{Temporal mode function of photon subtracted state}
\vspace{-2mm}
The quantum state from that two photons are subtracted belongs to a Fourier-transform limited mode whose temporal and spectral profiles are defined by the optical setup as well as the properties of the photon counters. We determine the temporal profile $f(t)$ in the course of our ensemble measurement.
We used the covariance matrix of the anti-squeezed quadrature $\hat{X}^Q$, following the approach of Ref. \cite{Morin2013}. Specifically, we sought a temporal mode that maximised the variance of the anti-squeezed quadrature, because this mode was expected to contain the two-photon subtracted state. Since we sampled the quadrature at $160$ discrete times $t_m$, the resulting covariance matrix  $C$ was a finite square matrix with $160$ columns and rows,
\[
C_{mn}=\langle \Delta \hat{X}^Q (t_m)\Delta \hat{X}^Q(t_n)\rangle.
\]
A reference measurement on ensemble of the vacuum state revealed that the covariance matrix of vacuum $D$ was not an identity matrix, and some correlations between nearby quadratures were present. This could be attributed to fast sampling 
of the quadratures together with finite bandwidth of the detector. The actually measured quadratures $\hat{X}^Q(t_j)$ were then given by the convolution of the input quadratures with temporal response function of the detector. 

The variance of the $\hat{X}$ quadrature of temporal mode $f(t)$ can be expressed as 
\[
\Delta^2 \hat X^Q_{\rm mode} = \frac{f(t)^T C f(t)}{f(t)^T D f(t)}. 
\]
Here $f(t)$ is treated as a column vector and the variance is properly normalised such that $\Delta^2 \hat X^Q_{\rm mode} = 1$ for vacuum state. The maximum achievable variance is given by the maximum eigenvalue of a renormalised covariance matrix
\[
\tilde{C}=D^{-1/2}C D^{-1/2},
\]
and the optimal mode function reads $D^{-1/2}f_1(t)$, where $f_1(t)$ is the eigenvector of $\tilde{C}$ that corresponds to the largest eigenvalue $\lambda_1$ of $\tilde{C}$. The 10 largest eigenvalues of $\tilde{C}$ are plotted in the inset of Fig.\,\ref{figS1}.
We can see that a single eigenvalue clearly stands out, indicating presence of a single well defined mode that can be associated with the two-photon subtracted state.

\begin{figure}[t!!!!!!!!!!!!!!]
\includegraphics[width=0.92\linewidth]{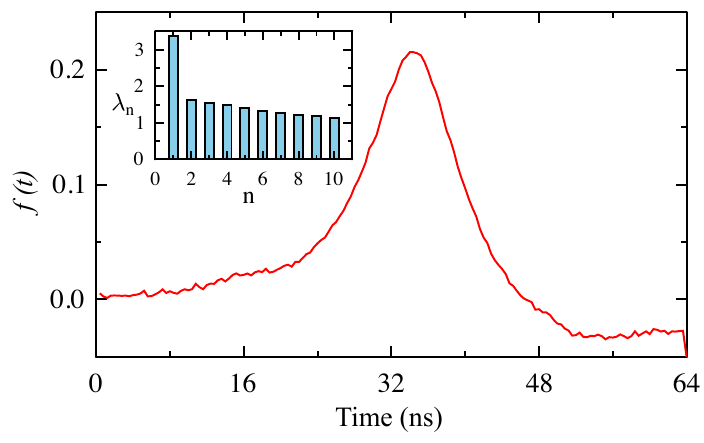}
\vspace{-3mm}
\caption{Optimal temporal mode profile $f(t) \approx f_1(t)$ determined form the measured covariance matrices. The inset shows the 10 largest eigenvalues $\lambda_n$ of the renormalised covariance matrix $\tilde{C}$ of the signal that contains the two-photon subtracted state.}
\label{figS1}
\end{figure}

When performing numerical calculations, we observed that the final multiplication of $f_1(t)$  by  $D^{-1/2}$ 
strongly enhanced the noise at high-frequency components of the mode function while the overall profile of the function almost did not change. 
We thus found that for our data we could omit the multiplication by $D^{-1/2}$ and use the mode profile $f_1(t)$ as an excellent approximation of the sought optimal mode function $f(t)$. 
In particular, the maximum eigenvalue of $\tilde{C}$ reads $\lambda_1 = 3.383$, and with the mode profile $f_1(t)$ we got $\Delta^2 X_{mode}^Q = 3.365$, which was very close. 
For comparison, if we avoided the renormalisation and considered a mode profile $f_C$ corresponding to the largest eigenvalue of the ``bare'' covariance matrix $C$, we got $\Delta^2 X_{mode}^Q = 3.220$. The mode function $f_1(t) \approx f(t)$ is plotted in Fig.\,\ref{figS1}. It was utilised in all subsequent data processing.

We used the same mode function for both balanced homodyne detectors of our 8-port device, just being shifted along the time axis to compensate for slightly different delays. Extra care was taken to design the two detectors such that they provided identical response. Naturally, the temporal dependence of the variance of anti-squeezed quadrature $\hat{X}^Q(t)$ exhibited much better signal-to noise ratio than 
the squeezed quadrature $\hat{Y}^Q(t)$. Therefore, the $\hat{X}^Q(t)$ data were used to construct the mode function. In fact, in the spectrum of the covariance matrix of the squeezed quadratures $\hat{Y}^Q(t_m)$ it was not possible to identify a single 
eigenvalue that would unambiguously correspond to the mode in the two-photon-subtracted state.
The measured quadratures of the two-photon subtracted state in mode $f_1(t)$ could be determined by integration over the measured $64$\,ns time window inside which $160$ samples were recorded,
\begin{eqnarray}
\hat{X}^Q_{\mathrm{mode}}&=&\sum_{m=1}^{160} f_+(t_m) \hat{X}^Q(t_m),  \nonumber \\
\hat{Y}^Q_{\mathrm{mode}}&=&\sum_{m=1}^{160} f_+(t_{m-d}) \hat{Y}^Q(t_m).
\end{eqnarray}
Here $d=2$ represents the identified offset between temporal delays in the two channels.

\section{Quantum state tomography}
\vspace{-2mm}
The 8-port homodyne detector sampled the Husimi Q-function, which represented a tomographically complete measurement. 
Here, we describe how we extracted the density matrix in Fock basis from the measured 8-port homodyne data.
We first constructed a histogram of the measured complex amplitudes $\beta_j$. We divided the phase space into rectangular bins with size $d=1/(8\sqrt{2})$ and centred at $\beta_{mn}=(m+in)d$, where $m,n\in \mathbb{Z}$. 
We counted the number $f_{mn}$ of measurement outcomes 
$\beta_j$ that felt inside the bin $(m,n)$ and we associated a POVM element 
\[
\Pi_{mn}=\frac{d^2}{\pi} |\beta_{mn}\rangle\langle \beta_{mn}|
\]
to each bin. 

We represented the operators and density matrices in Fock basis and introduced a cut-off at Fock state $N_{\mathrm{max}}=21$, which is sufficiently large to accommodate the investigated states with sufficient margin. 
We reconstructed the density matrix in Fock basis from the measured data $f_{mn}$ using an iterative Maximum-Likelihood reconstruction algorithm \cite{Jezek2003}, that seeks a state that is most likely to yield the observed experimental data. 
The algorithm naturally imposes all physical constraints on the reconstructed density matrix,  namely positive semi-definiteness and unit trace. The likelihood function is defined as
\begin{equation}
\mathcal{L}=\prod_{m,n} \left( p_{mn}\right)^{f_{mn}},
\end{equation}
where the product is taken over all nonzero $f_{mn}$, and
\begin{equation}
p_{mn}=\mathrm{Tr}[\hat{\Pi}_{mn}\hat{\rho}]
\end{equation}
is the theoretical probability of the measurement outcome $\hat{\Pi}_{mn}$.

It is convenient to work with the log-likelihood function $\ln \mathcal{L}$. The density matrix that maximises $\ln\mathcal{L}$ satisfies the extremal equation \cite{Hradil1997}
\begin{equation}
\hat{R}\hat{\rho}=\lambda\hat{\rho},
\label{rhoextremal}
\end{equation}
where 
\begin{equation}
\hat{R}=\sum_{m,n}\frac{f_{mn}}{p_{mn}}\hat{\Pi}_{mn} \, ,
\end{equation}
and  the Lagrange multiplier $\lambda$ accounts for the constraint $\mathrm{Tr}[\hat{\rho}]=1$. For a density matrix that maximises $\mathcal{L}$ and satisfies (\ref{rhoextremal}), we get $\lambda=\sum_{m,n}f_{mn}$.
Starting from a maximally mixed state in the truncated Fock space the density matrix that maximises $\mathcal{L}$ can be conveniently determined by repeated iterations of the following non-linear map that represents a symmetrised version of the extremal equation (\ref{rhoextremal}) \cite{Jezek2003},
\begin{equation}
\hat{\rho}\rightarrow \frac{\hat{R}\hat{\rho}\hat{R}^\dagger}{\mathrm{Tr}[\hat{R}\hat{\rho}\hat{R}^\dagger] }.
\end{equation}
Note that at each iteration step $\hat{R}$ depends on $\hat{\rho}$. 
This iterative reconstruction algorithm was applied to the original squeezed state, the two-photon subtracted state and a selected distilled state after one round of the gaussification protocol. In all cases a few hundred iterations proved sufficient to reconstruct the state.

The Wigner functions of the states plotted in Fig.\,4 of the main manuscript were obtained from the reconstructed density matrices in Fock basis. The reconstructed states were mixed, which could be mainly attributed to the overall losses in the setup and the imperfect two-photon subtraction. 
It is worth noting that the calculation of a Wigner function from the Husimi $Q$-functions amounts to a de-convolution, because
the $Q$-function can be expressed as a convolution of the Wigner function with a Gaussian function. 
From the point of view of the early linear quantum state tomography techniques, this deconvolution, when performed on noisy experimental data, 
may be difficult to implement and may lead to enhanced noise. 
However, the non-linear statistically motivated Maximum-Likelihood reconstruction is robust enough to perform
reliable quantum state reconstruction of non-classical states of light from the 8-port homodyne data.

\section{Gaussification}
\vspace{-2mm}
Here we review the calculation of the covariance matrix of the asymptotic Gaussian state as obtained by iterative Gaussification  \cite{Eisert2004,Campbell2012}. 
Let us first consider the original Gaussification procedure, where at each iteration two copies of the state interfere at a balanced beam splitter and the output port of the beam splitter that corresponds to constructive interference is projected onto vacuum. Upon success the output state at the other port is taken
 as an input for the next iteration of the protocol. A single step of this Gaussification protocol is described by a nonlinear map
\begin{equation}
\rho^{(j+1)}= \mathrm{Tr}_{2} \left[\left(\hat{U}_{\mathrm{BS}} \hat{\rho}^{(j)}\otimes \hat{\rho}^{(j)} \hat{U}_{\mathrm{BS}}^\dagger \right) \left (\hat{I}_1\otimes |0\rangle\langle 0|_2\right) \right],
\label{Gaussification}
\end{equation}
where $\mathrm{Tr}_2$ denotes the trace over the second mode and $\hat{U}_{\mathrm{BS}}$ denotes the unitary matrix of a balanced beam splitter. 
After the first iteration, the state $\hat{\rho}^{(1)}$ exhibits vanishing coherent displacement, $\langle \hat{X}\rangle=0$ and $\langle \hat{Y}\rangle=0$, and also $\rho^{(1)}_{1,0}=0$ due to destructive interference. 
If the nonlinear map (\ref{Gaussification}) converges, the asymptotic state $\hat{\rho}^{(\infty)}$ is Gaussian and its covariance matrix $\Gamma_G$ can be expressed as \cite{Eisert2004}
\begin{equation}
\Gamma_G=\Sigma^TB^{-1} \Sigma-I,
\label{GammaG}
\end{equation}
where $I$ is the $2\times 2$ identity matrix, $\Sigma$ denotes the symplectic form,
\vspace{-2mm}
\[
\Sigma=\left(
\begin{array}{cc}
0 & 1 \\
-1 & 0
\end{array}
\right),
\]
and the matrix $B$ can be expressed in terms of elements of normalised density matrix $\hat{\sigma}=\hat{\rho}^{(1)}/\rho_{0,0}^{(1)}$,
\begin{eqnarray}
B_{11}&=&\frac{1}{2}\left[1-\sigma_{1,1}+\sqrt{2}\mathrm{Re}(\sigma_{2,0})\right], \nonumber \\
B_{22}&=&\frac{1}{2}\left[1-\sigma_{1,1}-\sqrt{2}\mathrm{Re}(\sigma_{2,0})\right],  \nonumber \\
B_{12}&=&B_{21}= \frac{1}{\sqrt{2}}\mathrm{Im}(\sigma_{2,0}).  \nonumber 
\end{eqnarray}
The protocol will converge to a pure Gaussian state only if $\sigma_{1,1}=0$, otherwise the asymptotic Gaussian state with covariance matrix (\ref{GammaG}) will be mixed.

In our experiment we emulate the Gaussification protocol by processing the complex amplitudes $\beta$ measured with our 8-port homodyne detector. Consider two experimentally sampled complex amplitudes $\beta_1$ and $\beta_2$.  
The complex amplitudes corresponding to destructive- and constructive-interference output ports of the beam splitter read
\[
\beta_{-}=\frac{1}{\sqrt{2}} (\beta_1-\beta_2), \qquad  \beta_{+}=\frac{1}{\sqrt{2}} (\beta_1+\beta_2).
\]
The amplitude $\beta_{-}$ is accepted as an outcome of a single iteration of the Gaussification protocol if $\beta_{+}$ satisfies suitable acceptance condition. In experimental data processing, we impose a hard boundary $|\beta_{+}|^2\leq \bar{n}$. 
For theoretical analysis, it is more convenient to consider a Gaussian acceptance probability  
\[
P(\beta)=\exp\left(-\frac{|\beta|^2}{\bar{n}}\right).
\]
This acceptance condition gives rise to a modified Gaussification protocol, where one output mode of the beam splitter is effectively projected at the operator
\begin{eqnarray}
\hat{\Pi} &=& \frac{1}{\pi}\int_\beta P(\beta)|\beta\rangle\langle \beta| d^2\beta \nonumber \\
              &=& \frac{1}{\pi}  \int_\beta\exp\left(-\frac{|\beta|^2}{\bar{n}}\right) |\beta\rangle\langle \beta| d^2\beta,
\label{Pithermal}
\end{eqnarray}
and the formula (\ref{Gaussification}) changes to 
\begin{equation}
\rho^{(j+1)}= \mathrm{Tr}_{2} \left[\left(\hat{U}_{\mathrm{BS}} \hat{\rho}^{(j)}\otimes \hat{\rho}^{(j)} \hat{U}_{\mathrm{BS}}^\dagger \right) \left (\hat{I}_1\otimes \hat{\Pi}_2\right) \right],
\label{Ggeneralized}
\end{equation}
Up to normalisation, the expression  (\ref{Pithermal}) is  the density matrix of a thermal state with mean photon number $\bar{n}$ and we have
\begin{equation}
\hat{\Pi}= \sum_{n=0}^\infty \left( \frac{\bar{n}}{\bar{n}+1}\right)^{n+1}|n\rangle \langle n|.
\label{Pithermal}
\end{equation}
Convergence of generalised Gaussification protocols of the form (\ref{Ggeneralized}) was studied by Campbell and Eisert \cite{Campbell2012}, who derived analytical formula for the covariance matrix of the asymptotic Gaussian state for this case. One first defines a non-Hermitian operator 
\[
\hat{\sigma}=\frac{\hat{\rho}^{(1)}\hat{\Pi}}{\mathrm{Tr}\left[\hat{\rho}^{(1)}\hat{\Pi}\right]}
\]
that should exhibit zero coherent displacement, $\mathrm{Tr}[\hat{\sigma}\hat{X}]=\mathrm{Tr}[\hat{\sigma}\hat{Y}]=0$. Let $\Gamma_\sigma$ denote the generally complex-valued covariance matrix of $\hat{\sigma}$ and $\Gamma_{\Pi}$ the covariance matrix of the normalised operator $\hat{\Pi}/\mathrm{Tr}[\hat{\Pi}]$. 
Note that in the present case we have $\Gamma_{\Pi}=(1+2\bar{n})I$. The covariance matrix of the asymptotic Gaussian state can then be expressed as \cite{Campbell2012}
\begin{equation}
\Gamma_\infty=(\Gamma_\Pi -i\Sigma) (\Gamma_{\Pi}-\Gamma_{\sigma})^{-1}(\Gamma_{\Pi}+i\Sigma)-\Gamma_{\Pi}.
\label{Gammainfinity}
\end{equation}
The conditions for a (weak) convergence of the protocol are that the characteristic function of the operator $\hat{\sigma}$ satisfies $|\chi_\sigma(\vec{\xi})|\leq 1$ for all $\vec{\xi}$ and that the covariance matrix $\Gamma_\infty$ exists and is positive definite.

We now provide an alternative equivalent expression for the covariance matrix $\Gamma_{\infty}$ based on the re-interpretation of the Gaussification protocol (\ref{Ggeneralized}). Our approach avoids the use of non-Hermitian operators. We observe that a transmission of a quantum state $\hat{\rho}$ through a lossy quantum 
channel $\mathcal{L}$ with transmittance $T$ followed by projection onto vacuum corresponds to a generalised measurement on the original state $\hat{\rho}$ described by the operator 
\begin{equation}
\hat{\Pi}^\prime=\sum_{n=0}^\infty (1-T)^n |n\rangle\langle n|= \sum_{n=0}^\infty \left( \frac{\bar{n}}{\bar{n}+1}\right)^{n}|n\rangle \langle n|,
\label{Pilossy}
\end{equation}
where $T=1/(\bar{n}+1)$. Therefore, $\hat{\Pi}^\prime=\frac{\bar{n}+1}{\bar{n}}\hat{\Pi}$ and the operators (\ref{Pithermal}) and (\ref{Pilossy}) are identical up to a multiplicative constant. 
We next note that interference of two states at a beam splitter followed by transmission of each output state through a lossy channel $\mathcal{L}$ is equivalent to transmission of each input state through a lossy channel followed by interference at a beam splitter. 
Consequently, we can consider a modified input state $\hat{\rho}_{\mathcal{L}}^{(1)}=\mathcal{L}(\hat{\rho}^{(1)})$ and determine from Eq. (\ref{GammaG}) the covariance matrix $\Gamma_{G}$ for the original Gaussification protocol with projection onto vacuum.  
This covariance matrix must be interpreted as a covariance matrix of the asymptotic state transmitted through the lossy channel, $\Gamma_G=T\Gamma_\infty+(1-T)I$. This relation can be inverted and we get 
\begin{equation}
\Gamma_{\infty}= \frac{1}{T}\left[\Gamma_G-(1-T)I\right].
\label{Gammainfinityalternative}
\end{equation}
Numerical calculations confirm that the formulas (\ref{Gammainfinity}) and (\ref{Gammainfinityalternative}) yield the same covariance matrices. We take the reconstructed density matrix of the two-photon subtracted squeezed vacuum state and calculate the covariance matrix $\Gamma_\infty$ of the 
asymptotic state of the Gaussification protocol for various thresholds $\bar{n}$. We find that for our state the Gaussification protocol converges for $\bar{n}>0.3$. The distilled Gaussian state is mixed and close to this threshold the variance of the anti-squeezed quadrature becomes arbitrarily large, while the amount of distillable squeezing becomes limited, $\Delta^2 Y \gtrsim 0.35$. Therefore, it is reasonable to choose larger acceptance threshold $\bar{n}$ as it also increases the success rate of the protocol, c.f. Fig.\,5 in the main text.

\section{Photon subtraction augmented by coherent displacement}
\vspace{-2mm}
Here we discuss a more general photon subtraction operation, which is combined with a coherent displacement operation resulting in the conditional operation $\hat{a}+\delta$. This operation can be implemented either by coherently displacing the signal before or after the photon subtraction,
\begin{equation}
\hat{D}(-\delta) \hat{a}\hat{D}(\delta)=\hat{a}+\delta,
\end{equation}
or, more conveniently, by coherently displacing the tapped mode that is detected by the single photon detector \cite{Neergaard-Nielsen2010}. 
Here we consider a combination of two such `displaced photon subtractions' and choose the two coherent amplitudes such that the resulting operation preserves the parity of  Fock states,
\begin{equation}
\hat{M}=(\hat{a}+\delta)(\hat{a}-\delta)=\hat{a}^2-\delta^2.
\label{Moperator}
\end{equation}
Since $\delta$ is a complex amplitude, $\delta^2$ can be negative as well as positive. Application of  the generalised two-photon subtraction (\ref{Moperator}) to  pure squeezed vacuum state $|\psi(r)\rangle$ yields the non-normalised state
\begin{widetext}
\begin{equation}
|\psi_{2S}(r)\rangle= \frac{1}{\sqrt{\cosh(r)}}\sum_{n=0}^\infty\left[ (2n+1)\tanh r-\delta^2\right ]  (\tanh r)^n\frac{\sqrt{(2n)!}}{2^n n!}|2n\rangle.
\label{psi2Sgeneralized}
\end{equation}

The variances of the amplitude and phase quadratures of this state can be expressed as
\begin{eqnarray}
\Delta^2 \hat X &=& \, e^{2r} \left[1+4 \sinh^2 r \frac{2\sinh^2 r  + \cosh r \sinh r -\delta^2}{2\sinh^4 r+(\cosh r \sinh r-\delta^2)^2}\right] ,   \nonumber \\[2mm]
\Delta^2 \hat Y &=& e^{-2r} \left[1+4 \sinh^2 r \frac{2\sinh^2 r  - \cosh r \sinh r +\delta^2}{2\sinh^4 r+(\cosh r \sinh r-\delta^2)^2}\right].
\end{eqnarray}
\end{widetext}
The amplitude $\delta$ can be chosen to minimise the squeezed variance  of the two-photon subtracted state. Minimisation of  $\Delta^2 \hat Y$ with respect to $\delta^2$ yields
\[
\delta^2=\cosh r \sinh r-(2+\sqrt{6})\sinh^2 r.
\]
For this amplitude, the quadrature variances of the two-photon subtracted state (\ref{psi2Sgeneralized}) become 
\[
 \Delta^2 \hat X =\frac{7+2\sqrt{6}}{3+\sqrt{6}} e^{2r}, \qquad \Delta^2 \hat Y =\frac{3}{3+\sqrt{6}} e^{-2r}, 
\]
hence the optimised photon subtraction increases the squeezing by $2.6$ dB for arbitrary initial squeeze parameter $r$. This procedure can generate a state with finite squeezing from arbitrarily weakly squeezed input state, but the success probability of photon subtraction will decrease with decreasing $r$  and will scale as $\sinh^4 r$.

The squeezing of the photon subtracted state can be further increased by iterative Gaussification. The final squeeze parameter of the gaussified state  $r_{\rm G}$ can be expressed as 
\[
\tanh r_{\rm G}= \frac{3\tanh r -\delta^2}{\tanh r-\delta^2} \tanh r.
\]
This expression is meaningful and the gaussification converges only if $|\tanh r_{\rm G}|<1$. Note that any required squeezing $s>r$ is in principle achievable for any non-zero input squeezing $r$, if we set
\[
\delta^2=\frac{\tanh r_{\rm G} -3\tanh r}{\tanh r_{\rm G} -\tanh r} \tanh r.
\]
Therefore, arbitrary strong squeezing can be distilled from arbitrary weak initial squeezing by combination of the displacement-enhanced two-photon subtraction (\ref{Moperator}) followed by iterative Gaussification.


~\\
~\\
~\\
~\\

END OF SUPPLEMENTARY MATERIAL

\end{document}